\newcommand{\openone}{\leavevmode\hbox{\small1\normalsize\kern-.33em1}}
\documentclass[10pt]{iopart}
\usepackage{iopams,bm,cite,graphicx}

\begin{document}
\title{Geometrical approach to mutually unbiased bases}
\author{Andrei~B~Klimov$^1$, Jos\'e~L~Romero$^1$, Gunnar~Bj\"{o}rk$^2$
and Luis~L~S\'anchez-Soto$^3$}

\address{$^1$ Departamento de F\'{\i}sica,
Universidad de Guadalajara, 44420~Guadalajara,
Jalisco, Mexico}

\address{$^2$ School of Information and
Communication Technology,
Royal Institute of Technology (KTH), Electrum 229,
SE-164 40 Kista, Sweden}

\address{$^3$ Departamento de \'{O}ptica,
Facultad de F\'{\i}sica, Universidad Complutense,
28040~Madrid, Spain}
\date{\today}

\begin{abstract}
We propose a unifying phase-space approach to the construction of
mutually unbiased bases for a two-qubit system. It is based on an
explicit classification of the geometrical structures compatible
with the notion of unbiasedness. These consist of bundles of
discrete curves intersecting only at the origin and satisfying
certain additional properties. We also consider the feasible
transformations between different kinds of curves and show that they
correspond to local rotations around the Bloch-sphere principal
axes. We suggest how to generalize the method to systems in
dimensions that are powers of a prime.
\end{abstract}

\eqnobysec

\section{Introduction}

The notion of mutually unbiased bases (MUBs) emerged in the
seminal work of Schwinger~\cite{Schwinger60} and it has
turned into a cornerstone of the modern quantum information.
Indeed, MUBs play a central role in a proper understanding
of complementarity~\cite{Wootters87,Kraus87,Lawrence02,Chaturvedi02,Wootters04b},
as well as in approaching some relevant issues such as optimum
state reconstruction~\cite{Wootters89,Asplund01}, quantum
key distribution~\cite{Bechmann00,Cerf02}, quantum error
correction codes~\cite{Gottesman96,Calderbank97}, and
the mean king problem~\cite{Vaidman87,Englert01,Aravind03,Schulz03,Kimura06}.

For a $d$-dimensional system (also known as a qudit) it
has been found that the maximum number of MUBs cannot be
greater than $d+1$ and this limit is reached if $d=p$ is
prime~\cite{Ivanovic81} or power of prime, $d=p^{n}$~\cite{Calderbank97b}.
It was shown in reference~\cite{Ban02} that the
construction of MUBs is closely related to the
possibility of finding of $d+1$ disjoint classes,
each one having $d-1$ commuting operators, so that
the corresponding eigenstates form sets of MUBs. Since
then, different explicit constructions of MUBs in prime
power dimensions have been suggested in a number of
papers~\cite{Kla04,Lawrence04,Parthasarathy04,Pitt05,Durt05,Planat05,JPA05}.

The phase space of of a qudit can be seen as a
$d \times d$ lattice whose coordinates are elements
of the finite Galois field  $GF(d)$~\cite{Lidl86}. At
first sight, the use of elements of $GF(d)$ as coordinates
could be seen as an unnecessary complication, but
it proves to be an essential step: only by doing
this we can endow the phase-space grid with the same
geometric properties as the ordinary plane. There are
several possibilities for mapping quantum states onto
this phase space~\cite{Buot74,Galetti88,Cohendet88}.
However, special mention must be made of the elegant
approach developed by Wootters and coworkers in
references~\cite{Wootters04} and \cite{Gibbons04},
which has been used to define a discrete Wigner
function (see references~\cite{Paz05} and \cite{Durt06}
for picturing qubits in phase space). Any good assignment
of quantum states to lines is called a `quantum net''.
In fact, there is not a unique quantum net for a given
$d \times d$ phase space. However, one can manage to
construct lines and striations (sets of parallel lines)
in this phase space: after an arbitrary choice that does
not lead to anything fundamentally new, it turns out
that the orthogonal bases associated with each
striation are mutually unbiased.

In this paper, we proceed just in the opposite way. We
start by considering the geometrical structures in phase
space that are compatible with the notion of unbiasedness.
By taking the case of two identical two-dimensional systems
(i.e., two qubits) as the thread for our approach, we classify
these admissible structures into rays and curves (and the former
also in regular and exceptional, depending on the degeneracy).
To each bundle of curves, we associate a MUB, and we show how
these MUBs are related by local transformations that do not
change the corresponding entanglement properties. Finally,
we sketch how to extend this theory to higher (power of prime)
dimensions. We hope that this new method can seed light on
the structure of MUBs and can help to resolve some of the
open problems in this field. For example, all the MUB
structures in 8- and 16-dimensional Hilbert space are
known~\cite{PRA05}, but in the 16-dimensional case the
transformations to go from one structure to any other
are unknown and hitherto a method to find them (in any
space dimension) has been lacking. Our approach provides
a means to find such transformations in a systematic manner.

\section{Constructing a set of mutually unbiased bases}

When the space dimension $d = p^{n}$ is a power of a prime it is
natural to conceive the system as composed of $n$ constituents,
each of dimension $p$~\cite{Vourdas04}. We briefly summarize a
simple construction of MUBs for this case, according to the
method introduced in reference~\cite{JPA05}, although focusing
on the particular case of two-qubits. The main idea consists
in labeling both the states of the subsystems and the generators
of the generalized Pauli group (acting in the four-dimensional
Hilbert space) with elements of the finite field $GF(4)$, instead
of natural numbers. In particular, we shall denote as $ | \alpha
\rangle$ with $\alpha \in GF (4)$ an orthonormal basis in the
Hilbert space of the system. Operationally, the elements of the
basis can be labeled by powers of a primitive element (that is,
a root of the minimal irreducible polynomial, $\sigma^{2} +
\sigma + 1 = 0$), so that the basis reads
\begin{equation}
\{ |0 \rangle, \, |\sigma \rangle , \, | \sigma^{2} =
\sigma + 1 \rangle , \, | \sigma^{3} = 1 \rangle \} \, .
\end{equation}
These vectors are eigenvectors of the generalized position
operators $Z_{\beta}$
\begin{equation}
Z_{\beta} =  \sum_{\alpha \in GF(4) } \chi ( \alpha  \beta)
|\alpha \rangle \langle \alpha | \, ,
\end{equation}
where henceforth we assume $\alpha , \beta \in GF(4)$. Here
$\chi (\theta )$ is an additive character
\begin{equation}
\label{ch}
\chi ( \theta ) = \exp \left [ \frac{2\pi i}{p}
\tr ( \theta ) \right ] \, ,
\end{equation}
and the trace operation, which maps elements of $GF(4) $ onto
the prime field $GF(2) \simeq \mathbb{Z}_{2}$, is defined as
$\tr (\theta) = \theta + \theta^{2}$. The diagonal operators
$Z_{\beta}$ are conjugated to the generalized momentum
operators $X_{\beta}$
\begin{equation}
X_{\beta} = \sum_{\alpha \in GF(4)}
| \alpha + \beta \rangle \langle \alpha | \, ,
\end{equation}
precisely through the finite Fourier transform
\begin{equation}
\label{XZgf}
F \, X_{\beta} \, F^{\dagger} = Z_{\beta} \, ,
\end{equation}
with
\begin{equation}
F = \frac{1}{2} \sum_{\alpha, \beta \in GF(4)}
\chi ( \alpha \beta) \, | \alpha \rangle \langle \beta | \, .
\end{equation}
The operators $ \{Z_{\alpha}, X_{\beta}\}$ are the generators of
the generalized Pauli group
\begin{equation}
Z_{\alpha} X_{\beta} = \chi (\alpha \beta) \, X_{\beta} Z_{\alpha} \, .
\label{com}
\end{equation}
In consequence, we can form five sets of commuting operators
(which from now on will be called displacement operators) as
follows,
\begin{equation}
\label{set1}
\{ X_{\beta} \}, \{ Z_{\alpha} X_{\beta = \mu \alpha }\} \, ,
\end{equation}
with $\mu \in GF(4)$.  The displacement operators (\ref{set1})
can be factorized into products of powers of single-particle
operators $\sigma_{z}$ and $\sigma_{x}$, whose expression in
the standard basis of two-dimensional Hilbert space is
\begin{equation}
\label{primeZX}
\sigma_{z} = | 0 \rangle \langle 0 | - | 1 \rangle \langle 1 | \, ,
\qquad
\sigma_{x} = | 0 \rangle \langle 1 | + | 1 \rangle \langle 0 | \, .
\end{equation}
This factorization can be carried out by mapping  each element
of $GF(4)$ onto an ordered set of natural numbers~\cite{Gibbons04},
$\alpha \Leftrightarrow ( a_{1},a_{2}) $, where $a_{j}$ are
the coefficients of the expansion of $\alpha $ in a field basis
$\theta_j$
\begin{equation}
\alpha = a_{1} \theta_{1} + a_{2} \theta_{2} \, .
\end{equation}
A convenient field basis is that in which the finite Fourier
transform is factorized into a product of single-particle
Fourier operators. This is the so-called self-dual basis,
defined by the property $\tr ( \theta_{i} \theta_{j}) =
\delta _{ij}$. In our case the self-dual basis is
$(\sigma , \sigma^{2})$ and leads to the following
factorizations
\begin{equation}
Z_{\alpha} = \sigma_{z}^{a_1} \, \sigma_{z}^{a_2},
\qquad
X_{\beta} =  \sigma_{x}^{b_1} \, \sigma_{x}^{b_2},
\end{equation}
where $\alpha = a_{1} \sigma + a_{2} \sigma^{2}$ and $\beta =
b_{1} \sigma + b_{2}\sigma^{2}$. Using this factorization, one
can immediately check that, among the five MUBs that exist in
this case, three are factorable and two are maximally
entangled~\cite{Englert00}. Although the factorization of a
particular displacement operator depends on the choice of a
basis in the field, the global separability properties (i.e.,
the number of factorable and maximally entangled MUBs) is
basis independent. That is, any nonlocal unitary transformation
that yields only factorable or maximally entangled bases (i.e.,
a transformation from the Clifford group) will provide an isomorphic
set of MUBs with respect to the separability, except, perhaps, for
some trivial permutations. Nevertheless, this property holds only
for two qubits because for higher-dimensional cases more
complicated structures arise~\cite{PRA05}.

\section{Mapping the mutually unbiased bases onto phase space}

The problem of MUBs can be further clarified by an appropriate
representation in phase space, which is defined as a collection of
ordered points $(\alpha, \beta ) \in GF( 4) \times GF( 4) $. In this
finite phase space the operators  from the five sets (\ref{set1})
are labeled by points of rays (i.e., `straight´´ lines passing
through the origin). The vertical axis has $\alpha =0$ and the
horizontal axis has $\beta = 0$. For our case, we explicitly
have
\begin{eqnarray}
\label{set1a}
\beta =0 & \rightarrow & Z_{\sigma}, Z_{\sigma^2}, Z_{\sigma^3}
\nonumber \\
\beta = \alpha & \rightarrow & Z_{\sigma} X_{\sigma},
Z_{\sigma^{2}}X_{\sigma^2}, Z_{\sigma^3} X_{\sigma^3}
\nonumber \\
\beta = \sigma \alpha & \rightarrow & Z_{\sigma} X_{\sigma^2},
Z_{\sigma^2} X_{\sigma^3}, Z_{\sigma^3} X_{\sigma}   \\
\beta = \sigma^2 \alpha & \rightarrow & Z_{\sigma} X_{\sigma^3},
Z_{\sigma^2} X_{\sigma}, Z_{\sigma^3} X_{\sigma^2}
\nonumber \\
\alpha =0 & \rightarrow & X_{\sigma}, X_{\sigma^2}, X_{\sigma^3}
\nonumber
\end{eqnarray}
where the left column indicates the ray corresponding to the
operators appearing in the three rightmost columns. In the
factorized form, the set in (\ref{set1a}) can be expressed as
in table~1.

\begin{table}
\caption{Rays and their associated physical operators.}
\begin{indented}
\item[] \begin{tabular}{@{}lllll}
\br
Basis &  Ray &  \multicolumn{3}{c}{Factorized operators} \\
\mr
1 & $\beta = 0$ & $\sigma_{z} \openone$ & $\openone \sigma_{z}$
& $\sigma_{z} \sigma_{z}$ \\
2 & $\beta = \alpha $ & $\sigma_{y} \openone$ & $ \openone \sigma_{y}$
& $\sigma_{y}\sigma_{y}$ \\
3 & $\beta = \sigma \alpha$ & $\sigma_{z} \sigma_{x}$
& $\sigma_{x} \sigma_{y}$ & $\sigma_{y} \sigma _{z} $ \\
4 & $\beta = \sigma^{2} \alpha$ & $\sigma_{y} \sigma_{x}$
& $\sigma_{x} \sigma_{z}$ & $\sigma_{z} \sigma_{y}$ \\
5 & $\alpha = 0$ & $\sigma_{x} \openone$ & $\openone \sigma_{x}$
& $\sigma_{x} \sigma_{x}$ \\
\br
\end{tabular}
\end{indented}
\end{table}

\begin{figure}
\label{fig1}
\centering
\includegraphics[height=2.5cm]{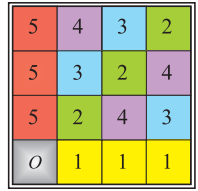}
\caption{Phase-space picture corresponding to the construction
in table~1.}
\end{figure}

In figure~1 we plot the phase-space representation of the
sets of operators in table~1. Each set has been
arbitrarily assigned to the number appearing in the left
column of the table. The sets of operators 1 and 5 define
the horizontal and the vertical axes, respectively, and they
lead, together with the operators associated to line 2, to
three separable bases (i.e., the three operators in each of
the first three rows commute for each of the two subsystems,
separately). In physical space, all these operators can be
associated  with rotations of each qubit around the $z$-, $x$-
and $y$-axis, respectively. Eigenstates of the operators
associated with the lines 3 and 4 form entangled bases (in
fact, their simultaneous eigenstates are all maximally
entangled states). The origin is labeled as $o$ and is the
common intersecting point of all the rays.

It is clear that under local transformations the factorable
and entangled MUBs preserve their separability properties.
Two natural questions thus arise in this respect: Is the
arrangement in table~1 and the corresponding geometrical
association with rays in phase-space unique? If this is
not the case, why do different arrangements always lead
to the same separability structure of MUBs?

\section{Curves in phase space}

To answer these questions we shall approach the problem from a
different perspective, namely, by determining all the possible
geometrical structures in phase space that correspond to MUBs. First
of all, let us observe that any ray can be defined in the
parametric form
\begin{equation}
\label{ray1}
\alpha (\kappa ) = \eta \kappa \, ,
\qquad
\beta (\kappa )= \zeta \kappa \, ,
\end{equation}
where $\eta, \zeta \in GF(4)$ are fixed while $\kappa \in GF(4)$
is a parameter that runs through all the field elements. The rays
(\ref{ray1}) can be seen as the simplest nonsingular (i.e., no
self-intersecting) Abelian substructures in phase space,
in the sense that
\begin{equation}
\label{ac}
\alpha ( \kappa + \kappa^\prime ) = \alpha (\kappa) +
\alpha ( \kappa^\prime ) \, ,
\qquad
\beta ( \kappa + \kappa^\prime ) = \beta (\kappa) +
\beta (\kappa^\prime) \, .
\end{equation}
However, the rays are not the only Abelian structures: it is
easy to see that the parametric curves (that obviously pass
through the origin)
\begin{equation}
\label{curve1}
\alpha ( \kappa ) = \mu_{0} \kappa + \mu_{1} \kappa^{2} \, ,
\qquad
\beta ( \kappa ) = \eta_{0} \kappa + \eta_{1} \kappa^{2} \, ,
\end{equation}
also satisfy the condition (\ref{ac}). If, in addition, we
impose
\begin{equation}
\label{Tr_cond}
\tr ( \alpha \beta^\prime ) =  \tr ( \alpha^\prime \beta ) \, ,
\end{equation}
where $\alpha^\prime = \alpha (\kappa^\prime )$ and $\beta^\prime =
\beta (\kappa^\prime)$, then the displacement operators associated
to the curves (\ref{curve1}) commute with each other and the
coefficients $\mu_{j}$ and $\eta_{j}$ must satisfy the following
restrictions (commutativity conditions)
\begin{equation}
\label{Eq:condition}
\mu_{1} \eta_{0} + (\mu_{1} \eta_{0})^{2} =
\mu_{0} \eta_{1} + (\mu_{0} \eta_{1})^{2} \, .
\end{equation}
All the possible Abelian curves satisfying condition (\ref{Eq:condition})
can be divided into two types:

a) regular curves
\begin{eqnarray}
\label{alphabeta}
\alpha\mathrm{-curves} & : & \quad
\alpha = \sigma \kappa \, ,
\quad
\beta = \eta \kappa + \sigma^{2} \kappa^{2} \, ,
\nonumber  \\
& & \\
\beta\mathrm{-curves} & : & \quad
\beta = \sigma \kappa \, ,
\quad
\alpha = \eta \kappa + \sigma^{2} \kappa^{2} \, .  \nonumber
\end{eqnarray}

b) exceptional curves
\begin{equation}
\label{exep}
\alpha = \mu (\kappa +\kappa^{2}) \, ,
\quad
\beta =\mu^{2} (\sigma \kappa + \sigma^{2} \kappa^{2}) \, .
\end{equation}

The regular curves are nondegenerate, in the sense that $\alpha $
or $\beta$ (or both) are not repeated in any set of four points
defining a curve. In other words, $\alpha $ or $\beta $ (or both)
take all the values in the field $GF(4)$. This allows us to write
down explicit relations between $\alpha $ and $\beta $ as follows
\begin{eqnarray}
\label{alpha1}
\alpha\mathrm{-curves} & : &
\quad
\beta = \eta \sigma^{2} \alpha + \alpha^{2} \, , \nonumber \\
& & \\
\beta\mathrm{-curves} & : &
\quad
\alpha = \eta \sigma^{2} \beta + \beta^{2} \, . \nonumber
\label{beta1}
\end{eqnarray}
By varying the parameter $\eta$ in the first of equations
(\ref{alpha1}) we can construct the $\alpha$-curves in
table~2, which show a different arrangement of operators
than (\ref{set1a}). Figure~2  shows the corresponding points of
$\alpha$-curves in phase space. Note, that we have completed
table~2 and figure~2 with the vertical ($X_{\sigma}, X_{\sigma^2},
X_{\sigma^3}$) axis. The factorization of operators in each
table (the self-dual basis is used for the representation
of operators in terms of Pauli matrices) is different from the
standard one in table~1. The curves marked as 3, 4 and 5 lead
now to factorable MUBs, while the ones marked as 1 and 2 lead
to maximally entangled bases.

The $\beta$-curves and the corresponding table can be obtained from
table~2 by exchanging $\alpha $ and $\beta $ (and correspondingly
$Z$ and $X$ operators) and is given in table~3. The phase-space
picture corresponding to table~3 is shown in figure~3 and can be
easily obtained from figure~2 by mirroring this figure about the
main diagonal. Observe that the curves $\beta =\alpha^{2}$ and
$\alpha =\beta^{2}$ then become identical, since this curve is
symmetric about the diagonal.

\begin{table}
\caption{$\alpha$-curves and their corresponding operators.}
\begin{indented}
\item[] \begin{tabular}{@{}llllll}
\br
Basis  &  $\alpha$-curves  &  Displacement operators &
\multicolumn{3}{c}{Factorized operators} \\
\mr
1 &  $\beta = \alpha^{2}$ & $Z_{\sigma^{2}} X_{\sigma},
Z_{\sigma^3} X_{\sigma^3}, Z_{\sigma} X_{\sigma^2}$ &
$\sigma_{x} \sigma_{z}$ & $\sigma_{y}\sigma_{y}$ &
$\sigma_{z}\sigma_{x}$ \\
2 &  $\beta = \alpha + \alpha^{2}$ & $Z_{\sigma^2} X_{\sigma^3},
Z_{\sigma^3}, Z_{\sigma} X_{\sigma^3}$ &
$\sigma_{x} \sigma_{y}$ & $\sigma_{z} \sigma_{z}$ &
$\sigma_{y}\sigma _{x}$ \\
3 & $\beta =\sigma \alpha + \alpha^{2}$ &
$Z_{\sigma^2} X_{\sigma^2}, Z_{\sigma^3} X_{\sigma^2}, Z_{\sigma}$ &
$\openone \sigma_{y}$ & $\sigma_{z} \sigma_{y}$ &
$\sigma_{z}\openone$ \\
4 & $\beta = \sigma^{2} \alpha + \alpha^{2}$ & $Z_{\sigma^2},
Z_{\sigma^3} X_{\sigma}, Z_{\sigma} X_{\sigma}$ &
$\openone \sigma_{z}$ & $\sigma_{y} \sigma_{z}$ &
$\sigma_{y}\openone$ \\
5 & $\alpha =0$ & $X_{\sigma}, X_{\sigma^2}, X_{\sigma^3}$ &
$\sigma_{x} \openone$ & $\openone \sigma_{x}$ & $\sigma_{x} \sigma_{x}$\\
\br
\end{tabular}
\end{indented}
\end{table}
\begin{figure}
\label{fig2}
\centering
\includegraphics[height=2.5cm]{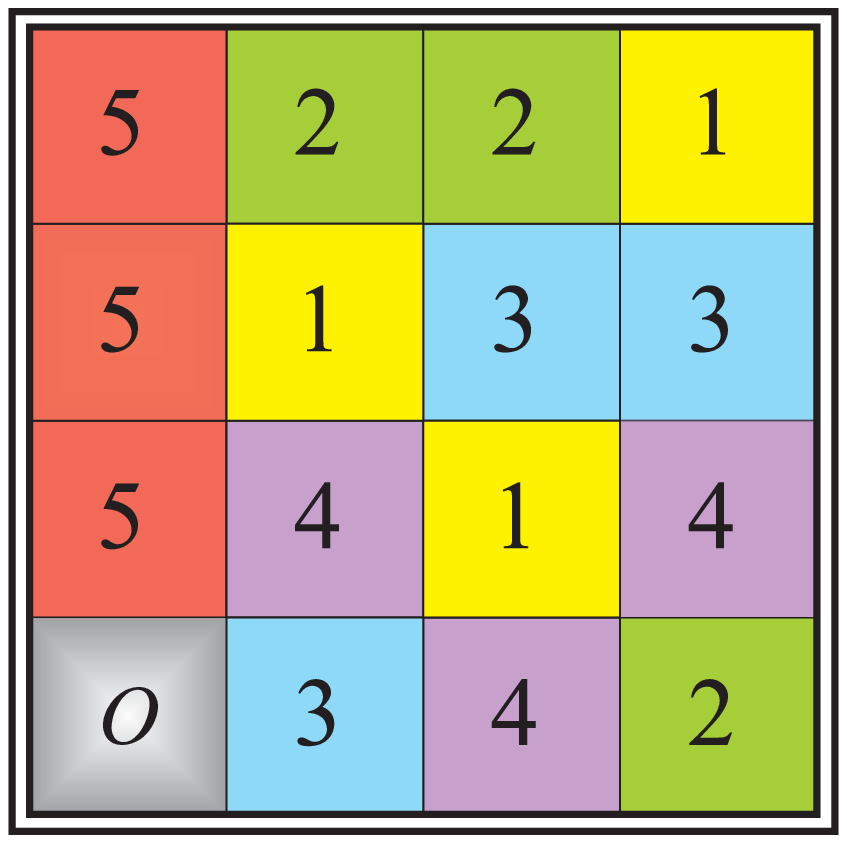}
\caption{Phase-space picture corresponding to the construction
in table 2.}
\end{figure}

\begin{table}[b]
\caption{Phase-space $\beta$-curves and their corresponding
operators.}
\begin{indented}
\item[] \begin{tabular}{@{}llllll}
\br Basis & $beta$-curves & Displacement operators &
\multicolumn{3}{c}{Factorized operators} \\
\mr
1 & $\alpha = \beta^{2}$ & $X_{\sigma^2} Z_{\sigma},
X_{\sigma^3} Z_{\sigma^3}, X_{\sigma} Z_{\sigma^2}$ &
$\sigma_{z} \sigma_{x}$ & $\sigma_{y} \sigma_{y}$ &
$\sigma_{x} \sigma_{z}$ \\
2 & $\alpha =\beta + \beta^{2}$ & $ X_{\sigma^2} Z_{\sigma^3},
X_{\sigma^3}, X_{\sigma} Z_{\sigma^3}$ &
$\sigma_{z} \sigma_{y}$ & $\sigma_{x} \sigma_{x}$ &
$\sigma_{y} \sigma_{z}$ \\
3 & $\alpha = \sigma \beta + \beta^{2}$ & $
X_{\sigma^2} Z_{\sigma^2}, X_{\sigma^3} Z_{\sigma^2}, X_{\sigma}$ &
$\openone \sigma_{y}$ & $\sigma_{x} \sigma_{y}$ & $\sigma_{x} \openone$ \\
4 & $\alpha = \sigma^{2} \beta + \beta^{2}$ &
$X_{\sigma^2}, X_{\sigma^3} Z_{\sigma}, X_{\sigma} Z_{\sigma}$ &
$\openone \sigma_{x}$ & $\sigma_{y} \sigma_{x}$ & $\sigma_{y} \openone$ \\
5 & $\beta =0$ & $Z_{\sigma}, Z_{\sigma^2}, Z_{\sigma^3}$  &
$\sigma_{z} \openone$ & $\openone \sigma_{z}$ & $\sigma_{z} \sigma_{z}$ \\
\br
\end{tabular}
\end{indented}
\end{table}
\begin{figure}[b]
\label{fig3}
\centering
\includegraphics[height=2.5cm]{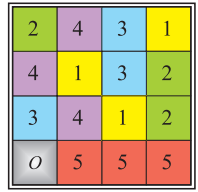}
\caption{Phase-space picture corresponding to the construction
in table 3.}
\end{figure}

It is worth noting that all the $\alpha$-curves, except $\beta =\alpha^{2}$,
are $\beta$-degenerate: the same value of $\beta $ corresponds to
different values of $\alpha $. Obviously, the analogous $\alpha$-degeneration
appears in the $\beta$-curves.

Exceptional curves (\ref{exep}) have quite a different structure. Now,
every point is doubly degenerate and can be obtained from equations
that relate powers of $\alpha $ and $\beta$:
\begin{equation}
\alpha^{2}=\mu \alpha \, ,
\qquad 
\beta^{2}=\mu^{2} \beta  \, .
\end{equation}
It is impossible to write an explicit nontrivial equation of the
form $f(\alpha ,\beta )=0$ for them. The existence of these curves
allows us to obtain interesting arrangements of MUB operators in
tables that do not contain any axis ($z$, $x$ or $y$). There are two
of such structures, shown in tables~4 and 5. As can be seen from
the rightmost column in both tables, the physical difference
between the two structures is that the two qubits are permuted
between them. The lines marked 2, 3 and 4 in both tables lead
to factorable MUBs, while the lines marked as 1 and 5 give
maximally entangled ones.

\begin{table}
\caption{Bundle consisting of two exceptional curves, one $\alpha $-curve, 
one $\beta $-curve, and a ray.}
\begin{indented}
\item[] \begin{tabular}{@{}llllll}
\br
Basis &  Curves and rays & Displacement operators &
\multicolumn{3}{c}{Factorized operators} \\
\mr
1 & $ \! \! \! \begin{array}{l}  \alpha = \kappa +\kappa^{2} \\
\beta =\sigma \kappa + \sigma^{2} \kappa^{2} \end{array} $  &
$X_{\sigma^3}, Z_{\sigma^3}, Z_{\sigma^3} X_{\sigma^3}$ &
$\sigma_{x} \sigma_{x}$ & $\sigma_{z} \sigma_{z}$ & $\sigma_{y} \sigma _{y}$ \\
2 & $ \! \! \! \begin{array}{l} \alpha = \sigma^2 (\kappa +\kappa^{2}) \\
\beta = \sigma^2 \kappa + \kappa^{2}  \end{array}  $ &
$X_{\sigma}, Z_{\sigma^2}, Z_{\sigma^2} X_{\sigma}$ &
$\sigma_{x} \openone$ & $\openone \sigma_{z}$ & $\sigma_{x} \sigma_{z}$ \\
3 & $\beta = \sigma \alpha + \alpha^{2}$ & $ Z_{\sigma^2} X_{\sigma^2},
Z_{\sigma^3} X_{\sigma^2}, Z_{\sigma}$ &
$\openone \sigma_{y}$ & $\sigma_{z} \sigma_{y}$ & $\sigma_{z} \openone$ \\
4 & $\alpha = \sigma^{2} \beta + \beta^{2}$ & $X_{\sigma^{2}},
Z_{\sigma} X_{\sigma^3}, Z_{\sigma} X_{\sigma}$ &
$\openone \sigma_{x}$ & $\sigma_{y} \sigma_{x}$ & $\sigma_{y} \openone$ \\
5 & $\beta = \sigma \alpha$ & $Z_{\sigma} X_{\sigma^2},
Z_{\sigma^2} X_{\sigma^3}, Z_{\sigma^3} X_{\sigma}$ &
$\sigma_{z} \sigma_{x}$ & $\sigma_{x}\sigma_{y}$ & $\sigma_{y}\sigma _{z}$ \\
\br
\end{tabular}
\end{indented}
\end{table}
\begin{figure}
\label{fig4}
\centering
\includegraphics[height=2.5cm]{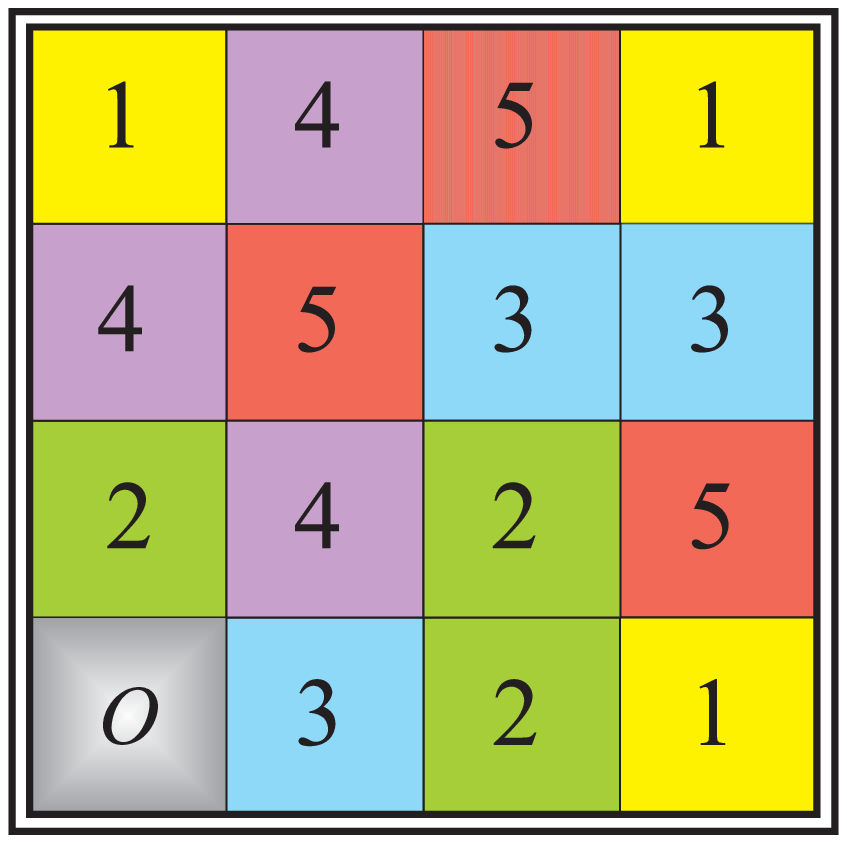}
\caption{Phase-space picture corresponding to the construction
in table 4.}
\end{figure}

\begin{table}[b]
\caption{Bundle consisting of two exceptional curves, 
one $\alpha $-curve, one $\beta $-curve, and a ray.}
\begin{indented}
\item[] \begin{tabular}{@{}llllll}
\br
Basis & Curves and rays & Displacement operators &
\multicolumn{3}{c}{Factorized operators}  \\
\mr
1 & $ \! \! \! \begin{array}{l}  \alpha =\kappa +\kappa^{2} \\
\beta =\sigma \kappa +\sigma^{2}\kappa^{2} \end{array}$ &
$X_{\sigma^3}, Z_{\sigma^3}, Z_{\sigma^3} X_{\sigma^3}$ &
$\sigma_{x} \sigma_{x}$ & $\sigma_{z} \sigma_{z}$ &
$\sigma_{y} \sigma_{y}$ \\
2 & $\! \! \! \begin{array}{l} \alpha = \sigma ( \kappa + \kappa^{2}) \\
\beta = \kappa + \sigma \kappa^{2} \end{array}$  &
$X_{\sigma^2}, Z_{\sigma}, Z_{\sigma} X_{\sigma^2}$ &
$\openone \sigma_{x}$ & $\sigma_{z} \openone$ & $ \sigma_{z} \sigma_{x}$ \\
3 & $\beta = \sigma^{2} \alpha + \alpha^{2}$ &
$Z_{\sigma^2},Z_{\sigma^3} X_{\sigma}, Z_{\sigma} X_{\sigma}$ &
$\openone \sigma_{z}$ & $\sigma_{y} \sigma_{z}$ & $\sigma_{y} \openone $ \\
4 & $\alpha = \sigma \beta + \beta^{2}$ &
$Z_{\sigma^2} X_{\sigma^2}, Z_{\sigma^2} X_{\sigma^3}, X_{\sigma}$ &
$\openone \sigma_{y}$ & $\sigma_{x} \sigma_{y}$ & $\sigma_{x} \openone$ \\
5 & $\beta = \sigma^{2} \alpha$ & $Z_{\sigma} X_{\sigma^3},
Z_{\sigma^2} X_{\sigma}, Z_{\sigma^3} X_{\sigma^2}$ &
$\sigma_{y} \sigma_{x}$ & $\sigma_{x} \sigma_{z}$ & $\sigma_{z} \sigma _{y}$ \\
\br
\end{tabular}
\end{indented}
\end{table}
\begin{figure}[b]
\label{fig5}
\centering
\includegraphics[height=2.5cm]{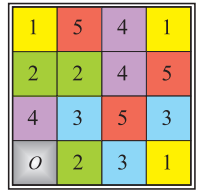}
\caption{Phase-space picture corresponding to the construction
in table 5.}
\end{figure}

Finally, there is a last table containing two exceptional curves
and a ray corresponding to the spin operators in the $y$-direction,
as it is shown in table~6.

To sum up, there exist fifteen different Abelian structures,
five rays and ten curves, which can be organized in six different
forms with the respect to MUBs. The existence of only six bundles
of mutually nonintersecting Abelian  nonsingular curves (i.e.,
different tables) also follows from the fact that the coset of
the full symplectic group, which preserves the commutation relations
(\ref{com}), on operations corresponding to nontrivial permutations
of columns and rows of (any) table [generated by the symplectic
group $Sp(2, GF(4))$], is precisely of order 6.

\begin{table}
\caption{Bundle consisting of two exceptional curves, 
one $\alpha $-curve, one $\beta $-curve, and a ray.}
\begin{indented}
\item[] \begin{tabular}{@{}llllll}
\br
Basis & Curves and rays & Displacement operators &
\multicolumn{3}{c}{Factorized operators}  \\
\mr
1 & $ \! \! \! \begin{array}{l}
\alpha = \sigma^{2} (\kappa + \kappa^{2}) \\
\beta =\sigma^{2}\kappa +\kappa^{2}
\end{array} $ &
$X_{\sigma^{2}}, Z_{\sigma}, Z_{\sigma} X_{\sigma^{2}}$ &
$\openone \sigma _{x}$ & $\sigma_{z} \openone$ &
$\sigma_{z} \sigma_{x}$ \\
2 & $\! \! \! \begin{array}{l} \alpha = \sigma ( \kappa +\kappa^{2}) \\
\beta =\sigma \kappa +\sigma \kappa^{2}
\end{array} $ &
$X_{\sigma}, Z_{\sigma^2}, Z_{\sigma^2} X_{\sigma}$ &
$\sigma_{x} \openone$ & $\openone \sigma_{z}$ & $\sigma_{x} \sigma_{z}$ \\
3 & $\beta = \alpha + \alpha^{2}$ & $Z_{\sigma^2} X_{\sigma^3},
Z_{\sigma^3}, Z_{\sigma} X_{\sigma^3}$ &
$\sigma_{y} \sigma_{x}$ & $\sigma_{z} \sigma_{z}$ & $\sigma_{x} \sigma_{y}$ \\
4 & $\alpha =\beta + \beta^{2}$ & $Z_{\sigma^3} X_{\sigma^2},
X_{\sigma^3}, Z_{\sigma^3} X_{\sigma}$ &
$\sigma_{z} \sigma_{y}$ & $\sigma_{x} \sigma_{x}$ & $\sigma_{y} \sigma_{z}$ \\
5 & $\beta = \alpha $ & $Z_{\sigma} X_{\sigma}, Z_{\sigma^2} X_{\sigma^2},
Z_{\sigma^3} X_{\sigma^3}$ &
$\sigma_{y} \openone$ & $\openone \sigma_{y}$ & $\sigma_{y} \sigma_{y}$ \\
\br
\end{tabular}
\end{indented}
\end{table}

\section{The effect of local transformations}

As we have noticed, different arrangements of operators in tables
(or bundling of phase-space curves) lead to the same separability
structure. To understand this point, let us study the effect of
local transformations. In other words, we wish to characterize
how a given curve changes when a local transformation is
applied to a set of operators labeled by points of this
curve.

To deal with such operations with curves, let us recall that
a generic displacement operator is factorized in the self-dual
basis as
\begin{equation}
Z_{\alpha} X_{\beta} = (\sigma_{z}^{a_1} \sigma_{x}^{b_1})
(\sigma_{z}^{a_2} \sigma_{x}^{b_2}) \equiv
( a_{1},b_{1} ) \otimes ( a_{2},b_{2} ) \, .
\end{equation}
It is clear that under local transformation (rotations by $\pi/2$
radians around the $z$-, $x$- or $y$-axes) applied to the $j$th
particle ($j=1,2$), the indices of the displacement operators
are transformed as follows:
\begin{eqnarray}
z\mathrm{-rotation} & : & \quad ( a_{j}, b_{j} ) \rightarrow
(a_{j} + b_{j}, b_{j}) \, , \nonumber \\
x\mathrm{-rotation} & :& \quad ( a_{j}, b_{j} ) \rightarrow
(a_{j}, b_{j} + a_{j}) \, , \\
y\mathrm{-rotation} & :& \quad ( a_{j},b_{j} ) \rightarrow
(a_{j} + a_{j} + b_{j}, b_{j} + a_{j}+ b_{j}) = ( b_{j},a_{j} ) \, .
\nonumber
\end{eqnarray}
To give a concrete example, suppose we consider a $z$-axis rotation.
The operator $\sigma_{z}$, corresponding to $(a_j=1, b_j=0)$, is
transformed into $(a_j = 1 + 0 = 1,  b_j=0)$; i.e., into itself,
while, e.g., the operator $\sigma_{x}$, corresponding to $(a_j=0,
b_j=1)$, is mapped onto $(a_j = 0 + 1 = 1, b_j=1)$, which coincides
with $\sigma_{y}$. In the same way $\sigma_{y}$ is mapped onto
$\sigma_{x}$, while the identity ($a_j=0$, $b_j=0$) is mapped
onto itself.

In terms of field elements these transformations read
\begin{eqnarray}
z\mathrm{-rotation} & : & \quad
\begin{array}{l} \alpha \rightarrow  \alpha +
\theta_{j} \tr( \beta \theta _{j}) , \\
\beta \rightarrow  \beta ,
\end{array}  \nonumber \\
x\mathrm{-rotation} & : & \quad
\begin{array}{l} \alpha \rightarrow  \alpha , \\
\beta \rightarrow   \beta + \theta_{j} \tr( \alpha \theta_{j}) ,
\end{array} \\
y\mathrm{-rotation} & : & \quad
\begin{array}{l}  \alpha \rightarrow
\alpha + \theta_{j} \tr [ ( \alpha + \beta ) \theta_{j} ] , \\
\beta \rightarrow  \beta + \theta_{j} \tr [ (\alpha + \beta ) \theta _{j} ] .
\end{array} \nonumber
\end{eqnarray}
In particular, applying the above transformations to a
ray (\ref{ray1}) we get
\begin{eqnarray}
\label{x}
z\mathrm{-rotation} & : & \quad
\begin{array}{l} \alpha \rightarrow (\eta + \zeta \theta_{j})
\kappa + \kappa^{2} \zeta^{2}, \\
\beta \rightarrow  \beta = \zeta \kappa ,
\end{array} \nonumber  \\
x\mathrm{-rotation} & : & \quad
\begin{array}{l}  \alpha \rightarrow \alpha =  \eta \kappa , \\
\beta \rightarrow  ( \zeta + \eta \theta_{j}) \kappa +
\kappa^{2} \eta^{2} , \end{array} \\
y\mathrm{-rotation} & : & \quad
\begin{array}{l} \alpha \rightarrow ( \eta +
\zeta \theta_{j} + \eta \theta_{j} ) \kappa +
\kappa^{2} (\zeta + \eta )^{2}, \\
\beta \rightarrow ( \zeta + \zeta \theta_{j} +
\eta \theta_{j}) \kappa + \kappa^{2}(\zeta + \eta )^{2},
\end{array}    \nonumber
\end{eqnarray}
which are explicitly nonlinear operations.

Note that the $z$- and $x$-transformations produce regular
curves starting from a ray
\begin{eqnarray}
z\mathrm{-rotation} & : & \quad
\alpha = \eta \zeta^{-1} \beta
\rightarrow ( \eta \zeta^{-1} + \theta_{j} ) \beta + \beta^{2} ,
\nonumber \\
& & \\
x\mathrm{-rotation} & : & \quad
\beta = \eta^{-1} \zeta \alpha \rightarrow  ( \eta^{-1} \zeta +
\theta_{j} ) \alpha + \alpha^{2}. \nonumber
\end{eqnarray}
Meanwhile, the $y$-rotation may lead to an exceptional curve
(as it happens when we start with the horizontal or the
vertical axes, $\zeta =0$ or $\eta =0$).

An important result to stress is that it is possible to
obtain all the curves of the form (\ref{alphabeta}) and
(\ref{exep}) from the rays after some (nonlinear) operations
(\ref{x}), corresponding to local transformations.
The families of such transformations are the following:

I. The rays and curves corresponding to factorable basis
can be obtained from a single ray $\alpha =0, \beta =\sigma^{2}
\kappa $ (vertical axis) as shown in table~7 (left).

II. The rays and curves corresponding to nonfactorable basis can be
obtained from the ray $\alpha =\sigma \kappa ,\beta =\sigma^{2}\kappa $
($\beta =\sigma \alpha )$ as shown in table~7 (right).

\begin{table}
\caption{Curves and their corresponding transformations from
$\alpha =0, \beta = \sigma^{2} \kappa $ (left) and
$ \beta = \sigma \alpha$ (right). The $x$-, $y$ and
$z$-rotations are indicated as $x$, $y$ and $z$,
respectively.}
\begin{indented}
\item[] \begin{tabular}{@{}ll||ll}
\br
Curve (ray) & Transformation & Curve (ray) & Transformation \\
\mr
$ \! \! \!  \begin{array}{l}
\alpha = \sigma^{2} ( \kappa + \kappa^{2}) \\
\beta =\sigma^{2}\kappa + \kappa^{2}
\end{array} $ &
$\openone \otimes y$ & $ \! \! \!  \begin{array}{l}
\alpha = \kappa + \kappa^{2} \\
\beta =\sigma \kappa + \sigma^{2} \kappa^{2}
\end{array}$ & $z\otimes y$ \\
$ \! \! \!  \begin{array}{l}
\alpha =\sigma ( \kappa +\kappa^{2} ) \\
\beta =\sigma \kappa +\sigma \kappa^{2}
\end{array} $ & $y \otimes \openone$ &
$\beta =\sigma^{2}\alpha$ & $x\otimes x$ \\
$\beta =0$ & $y \otimes y$ &
$\beta =\alpha^{2}$ & $ \openone\otimes x$ \\
$\beta =\alpha$ & $z \otimes z$ &
$\beta =\alpha +\alpha^{2}$ & $\openone \otimes y$ \\
$\beta =\sigma \alpha +\alpha^{2}$ & $y \otimes z$ &
$\alpha =\beta +\beta^{2}$ & $y \otimes \openone$   \\
$\beta =\sigma^{2}\alpha + \alpha^{2}$ & $z \otimes y$ & &  \\
$\alpha =\sigma \beta + \beta^{2}$ & $\openone \otimes z$ & &  \\
$\alpha =\sigma^{2}\beta + \beta^{2}$ & $z \otimes \openone$ & & \\
\br
\end{tabular}
\end{indented}
\end{table}

This means that all the different tables can be generated from
the standard one, given in table~1, by applying only local
transformations that do not change the factorization
properties of the MUBs. So, tables 2 to 6 are obtained from
table~1 from the transformations given in table~8.

The full set of striations for each bundle of curves (each table)
is obtained by constructing ``parallel curves'' in the bundle
in an obvious way:
\begin{equation}
\alpha_{\lambda} ( \kappa ) = \mu_{0} \kappa +
\mu_{1} \kappa^{2},
\qquad
\beta_{\lambda} ( \kappa ) =  \eta_{0} \kappa +
\eta_{1} \kappa^{2} + \lambda ,
\end{equation}
with $\lambda \in GF(4)$. It is clear that no ($\alpha_{\lambda}
(\kappa ), \beta_{\lambda} ( \kappa)$) curve intersects the
curve ($\alpha_{\lambda^\prime} (\kappa ), \beta_{\lambda^\prime}
(\kappa )$) for $\lambda \neq \lambda^{\prime }$.

\section{Extension to larger spaces}

The relation between Abelian curves in discrete phase space
and different systems of MUBs can be extended to higher
(power of prime) dimensions. For the most interesting
$n$-qubit case, a generic Abelian curve (\ref{ac}) has
the following parametric from
\begin{equation}
\alpha ( \kappa ) = \sum_{m=0}^{n-1} \mu_{m} \kappa^{2^{m}} \, ,
\qquad
\beta ( \kappa ) = \sum_{m=0}^{n-1} \eta_{m} \kappa^{2^{m}} \, ,
\end{equation}
with $\mu_{m},  \eta _{m}, \kappa \in GF(2^{n})$,
and the commutativity condition takes now the
invariant form
\begin{equation}
\label{Tr_n}
\sum_{m \neq k} \tr( \mu_{m} \mathbf{\eta}_{k} )=  0 \, .
\end{equation}
The simplest example of such curves are obviously the rays,
parametrically defined as in Eq.~(\ref{ray1}), where the
conditions (\ref{Tr_n}) are trivially satisfied. Imposing
the nonintersecting condition we can, in principle, get all
the possible bundles of commutative curves. Nevertheless,
in higher dimensions it is impossible to obtain all the curves
from the rays by local transformations. This leads to the
existence of different nontrivial bundles of nonintersecting
curves, and consequently to MUBs with different types of
factorization~\cite{Lawrence04,PRA05}.

The problem of classification of bundles of mutually
nonintersecting, nonsingular Abelian curves and its relation to the
problem of MUBs in higher dimensions, and in particular the
transformation relations between different MUB structures, will be
considered elsewhere.

\begin{table}
\caption{Transformation operators converting table 1 into each
one of the tables indicated in the left column. Again, $x$-,
$y$- and $z$-rotations are indicated as $x$, $y$ and $z$,
respectively.}
\begin{indented}
\item[] \begin{tabular}{@{}ll}
\br
Table & Transformation \\
\mr
2 & $x\otimes \openone$ \\
3 & $\openone\otimes z$ \\
4 & $y\otimes z$ \\
5 & $y\otimes x$ \\
6 & $\openone\otimes y$ \\
\br
\end{tabular}
\end{indented}
\end{table}

\section{Conclusions}

A new MUB construction has been worked out, with special emphasis in
the two-qubit case. Its essential ingredient is a mapping between
displacement operators, physical spin-1/2 operators and discrete
phase-space curves. In phase space any nonsingular bundle of
curves that fills every point and has only one common intersecting
point (here taken to be the origin) will map onto a MUB. The
corresponding displacement operators can be obtained from these
phase-space curves.

For the two-qubit case, we have derived all the admissible
curves and classified them into rays and curves (regular
and exceptional, depending on degeneracy). In total, six
different bundles can be constructed from the set of five
rays and ten curves. We have also shown how the six tables
representing sets of MUBs are related by local transformations,
i.e., physical rotations around the $x$-, $y$- and $z$-axes.
It is obvious that such rotations will not change the MUBs
entanglement properties.

A Wigner function can also be associated to each phase-space
structure. Although we have not pursued this topic in the
paper, it is straightforward to use any of the phase-space
structures and follow the algorithm described in
reference~\cite{Gibbons04} (although in that paper
the construction applies only to rays) to obtain such
a function~\cite{JOPB07}.

It is also formally straightforward to extend the method to any
Hilbert space whose dimension is a power of a prime. However,
only in the bipartite case one will find that all structures
are related through local transformations. Already in the
tripartite case different classes of entanglement
exist~\cite{PRA05}, and consequently some MUB structures are
related through nonlocal (entangling) transformations. The
extention of the present method provides a systematic way
to find these transformations.

\section{Acknowledgements}

This work was supported by the Grant 45704 of Consejo Nacional
de Ciencia y Tecnologia (CONACyT), Mexico, the Swedish Foundation
for International Cooperation in Research and Higher Education
(STINT), the Swedish Research Council (VR), the Swedish Foundation
for Strategic Research (SSF), and the Spanish Research Directorate
(DGI), Grant FIS2005-0671.

\end{document}